\documentclass[12pt]{article}
\usepackage{epsfig}
\begin{document}
%set bwid 6; set fwid 6; set hwid 6; set pwid 6; set xwid 8; set ywid 6;
%igset lwid 6

\begin {center}
{\bf The covalent bond in Particle Spectroscopy}
\vskip 5mm
{D.\ V.\ Bugg\footnote{email: d.bugg@rl.ac.uk} \\
{\normalsize\it Queen Mary, University of London, London E1\,4NS, UK}
\\[3mm]}
\end {center}
\date{\today}

\begin{abstract}
\noindent
It is proposed that meson resonances are linear combinations of $q\bar
q$ and meson-meson (MM); baryon resonances are combinations of $qqq$
and meson-baryon (MB).
Mixing between these combinations arises via decays of confined states
to meson-meson or meson-baryon. There is a precise analogy with the
covalent bond in molecular physics; it helps to visualise what is
happening physically.
One eigenstate is lowered by the mixing;
the other moves up and normally increases in width.
Cusps arise at thresholds.
At sharp thresholds due to S-wave 2-particle decays, these
cusps play a conspicuous role in many sets of data.

\vskip 2mm

\end{abstract}

\section {Principles}
The Hamiltonian for a $q\bar q$ state decaying to meson-meson obeys
%Eq 1
\begin {equation}
H \Psi  = \left(
\begin {array} {cc}
H_{11} & V \\
V & H_{22}
\end {array}
\right) \Psi;
\end{equation}
$H_{11}$ describes short-range configurations
and $H_{22}$ refers to ingoing and outgoing states; $V$ accounts for
the coupling between them due to $s$-channel decays.
Then $H_{22}$ describes $s$-channel loop diagrams and also $t$ and 
$u$-channel exchanges.
Several authors adopt this principle, notably Jaffe
\cite {Jaffe};
he gives the equations and discusses many
of the implications.

The central premise of the present paper is that both $H_{11}$ and
$H_{22}$ play essential roles in all cases.
This is different from approaches where only one of the two
components in the Hamiltonian contributes, for example the approach
based on four-quark mesonic states.
It requires a review of all points of view and an appeal to data to
support the contention that both short and long-range configurations
are needed.

The wave function $\Psi$ is a linear combination of $q\bar q$ (or
$qqq$) and unconfined MM (or MB).
The crucial point is that two attractive components $H_{11}$ and
$H_{22}$ lower the eigenstate via the mixing.
This is a purely quantum mechanical effect.
There is a direct analogy with the covalent bond in chemistry.
The solution of Eq. (1) is
given by the Breit-Rabi equation:
%Eq 2
\begin {equation} E = (E_1 + E_2)/2 \pm \sqrt {(E_1 - E_2)^2 + |V|^2},
\end {equation}
where $E_1$ and $E_2$ are eigenvalues of separate $H_{11}$ and
$H_{22}$.
One linear combination  is pulled down in energy.

In chemistry, $H$ is in principle known exactly.
The discussion of the hydrogen molecule (and more complex ones) is
given in many textbooks on Physical Chemistry, for example the one of
Atkins \cite {Atkins}.
In the valence bond
approximation, $\Psi$ is described by two molecular ions, equivalent to
$q\bar q$ and $MM$, with a pair of electrons concentrated between the
ions in the ground state or repelled from this region in the upper
state.
Two key points are that (a) the extension of the wave function into the
overlap region lowers momentum, hence kinetic energy, (b) the whole
system shrinks slightly and the binding of electrons to both nuclei
increases.
This is the source of binding.
The analogue in Particle Physics is that the $MM$ wave function for the
ground state is sucked into the region of overlap and repelled
from it in the upper state, hence affecting zero-point energy.
The second effect is that the increased binding draws the quarks
slightly down the Coulombic part of the QCD potential, shrinking
the radius of the state.

There is, however, a further vital element in Particle Physics.
Resonances such as $f_0(980)$ are attracted to thresholds.
For $f_0(980)$, the amplitude for $\pi \pi \to KK$ is given to first
approximation by the Flatt\' e formula \cite {Flatte}:
%Eq 3
\begin {equation}
A_{12} = T_{12}\sqrt {\rho _1 \rho _2} =
\frac {G_1G_2 \sqrt {\rho _1 \rho _2 }}
{M^2 - s - i[G_1^2 \rho _1(s) + G_2^2 \rho _2(s)]}
=\frac {N(s)}{D(s)}
\end {equation}
where phase space is factored out of $T$.
Here $G_i = g_i F_i(s)$ and $g_i$ are coupling constants, $F_i$ are
form factors.
Writing $D(s) = M^2 - s - i\Pi (s)$, a more exact form for $D(s)$ is
$M^2 - s - Re \, \Pi (s) - i\Pi (s)$, where
%Eq 4
\begin {equation} Re \, \Pi _{KK}(s) =
\frac {1}{\pi} \rm {P} \int _{4m^2_K}
^\infty ds' \, \frac {G^2_{KK}(s') \rho _{KK}(s')} {(s' - s)}.
\end {equation}

Fig. 1 illustrates $Re \, \Pi (s)$ for $f_0(980)$ using
$F_{KK} = \exp (-3k^2_{KK})$, where $k_{KK}$ is $KK$ momentum in GeV/c.
$Re \, \Pi$ acts as an effective attraction \cite {Sync}.
Parameters of $f_0(980)$ are known.
Ref. \cite {Sync} gives tables of pole positions
when $M$, $g_1$ and $g_2$ are varied.
If $M$ is as low as  500 MeV, there is still a pole at
$806 - i78$ MeV;
for $M$ in the range 850--1100 MeV, there is a pole within 23 MeV
of the $KK$ threshold.
The moral is that a strong threshold can move a resonance a
surprisingly long way.
For $f_0(980)$, $g^2_{KK} \sim 0.7$ GeV$^2$.
%Fig. 1
\begin{figure}[htb]
\begin{center} \vskip -12mm
\epsfig{file=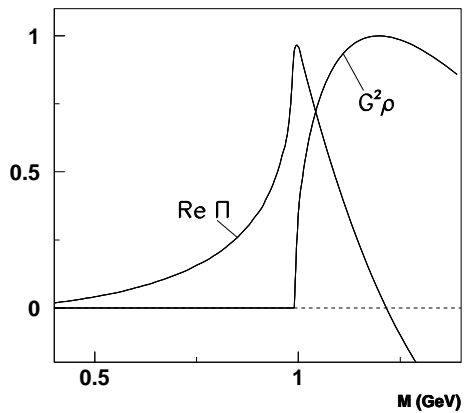,width=8cm}
\vskip -6mm
\caption
{$Re \, \Pi _{KK}(s)$ and $G^2_{KK}\rho _{KK}(s)$ for $f_0(980)$,
normalised to 1 at the peak of $G^2_{KK}\rho _{KK}$.}
\end{center}
\end{figure}

Since this article is written as much for experimentalists as
theorists, it is worth pausing briefly to describe what lies
behind Eq. (4).
There is a wide class of differential equations which obey the
principle of analyticity.
We do not know the exact equation, but in all of these equations,
real and imaginary parts of the amplitude are related.
It is possible to do experiments which measure the amplitude for
$\pi\pi \to KK$ at all energies.
Eq. (4) relates real and imaginary parts of the amplitude.
Another way of looking at this is that a $KK$ pair at one value
of $s$ can change its energy for as long as the uncertainty
principle allows and become `virtual' particles which explore
the whole range of $s$.
For this brief instant of time, the $KK$ pair describes a virtual
loop to other values of $s$, a loop diagram.
Evaluating Eq. (4)  is equivalent to evaluating
loop diagrams to all orders or solving the Bethe-Salpeter equation.
Confined $q\bar q$ configurations do decay and the logic behind
this paper is that we must therefore allow for effects due to the
meson pairs into which they decay, and the associated singularities
in $t$ and $u$ channels.

Barnes and Swanson \cite {Barnes} consider meson loops due to $D$,
$D^*$, $D_s$ and $D_s^*$ meson pairs.
For 1S, 1P and 2P charmonium states, they find large mass shifts which
may be `hidden' in the valence quark model by a change of parameters.
The important conclusion from their work is that two-meson continuum
components of charmonium states can be quite large, with the result
that the nieve constituent quark model may be misleading, particularly
near the thresholds of opening channels.
Lee et al. \cite {Lee} give an illuminating discussion of $X(3872)$,
discussed below in Section 2.1.

Eq. (4) has two virtues.
Firstly, it is easily evaluated, secondly it illustrates graphically
the effect of the form factor $F$.
Re $\Pi (s)$ goes negative close to the peak of $G^2 \rho$ and
subsequently has a minimum at $\sim 1.7$ GeV; thereafter it slowly
rises to 0 as $M \to \infty$.
Here an exponential form factor is used: $F = \exp (-3k^2)$, with
$k$ the centre of mass kaon momentum in GeV/c.

A difficulty at present is that the form factor is not known precisely.
Some authors use a different formula for the form factor.
The usual Flatt\' e formula, Eq. (3), serves as an approximate fitting
function where $M$ and $g^2$ are fitted empirically. A second point is
that the cusp changes slope abruptly at the $KK$ threshold.
Analysis of data then requires a precise knowledge of experimental
resolution if the cusp is included in the fit.
This is illustrated for $a_0(980)$ in  Ref. \cite {a0980}, where
Crystal Barrel data on $\bar pp \to \eta \pi ^0 \pi ^0$ are fitted
including the cusp.
The mass resolution (9.5 MeV) is known accurately in this case, but seriously
smears out the cusp.
A further detail is that separate thresholds for $K^+K^-$ and
$K^0 \bar K^0$ are not used in Fig. 1, for simplicity;
these two thresholds may be taken into account straightforwardly
\cite {Achasov}.

Oset and collaborators demonstrate in a series of papers that
resonances may be understood as `dynamically driven states'
\cite {Gamermann} \cite {Geng} \cite {Gonzalez} \cite {Geng2}
\cite {Molina} \cite {Sarkar} \cite {Molina2} \cite {Oset} due to
$s$, $t$ and $u$-channel exchanges.
They take the view that $\bar qq$ and $qqq$ components are not needed
at all in these cases.
However, the well known $J/\Psi$, $^3P_0(3415)$,
$^3P_1(3510)$, $^3P_2(3556)$, $\Psi (2S)$ and $\Psi (3770)$ are
interpreted naturally as $c\bar c$ states (with tiny admixtures of the
mesonic states to which they decay). Therefore it is logical to included
the $q\bar q$ component for all other resonances unless there is a good
reason why not.

How is it that Oset et al. are able to reproduce known resonances
(approximately) with meson exchanges alone?
They use S-wave form factors which are adjusted to get one predicted
resonance of each paper at the right mass.
Resulting amplitudes are strong.
The form factors may be mocking up short range $q\bar q$ components.
The importance of the work of Oset et al. is the demonstration that
components derived from meson loops are large, and should
be taken into account by theorists.
However, one should not be deterred from invoking $q\bar q$ and $qqq$
components to get all resonances with their correct masses and widths.

There is little evidence for resonance decays to $I=2$ $\pi \pi$
pairs or $I = 3/2~K\pi$, where interactions are repulsive
\cite {kappa}.
The commonly observed SU(3) octets and decuplets are those whose
decays do {\it not} lead to such repulsive final states.
Higher representions such as {\bf 27}, {\bf 10} and $\bf {\bar {10}}$
{\it do} lead to such repulsive final states.
The natural interpretation is that repulsive final states actively
inhibit formation of representations higher than octets and decuplets.
The Variational Principle arranges that the configurations which are produced
are those where repulsive final states are suppressed.

An approach which has recently been popular is to suppose that
`molecules' are formed from $\bar q \bar q q q$ configurations
\cite {Maiani} \cite {Braaten}\cite {Hanhart}  \cite {Alvarez}
\cite {Martinez}.
The well known problem with this approach is why so few tetraquarks
have been observed.
Vijande et al. \cite {Vijande} throw light on this issue.
They study the stability of pure $c\bar c n \bar n$ and
$cc \bar n \bar n$ states excluding diquark interactions.
They find that all 12 $c\bar c n\bar n$ states with $J = 0$, 1 or 2
are unstable.
The calculation points to the conclusion that such molecules are
rare unless either (i) there are attractive diquark interactions, or
(ii) coupling to meson-meson final states contributes, as proposed
here.

Jaffe has proposed \cite {Jaffe2} that $\sigma$, $\kappa$, $a_0(980)$
and $f_0(980)$  are colourless 4-quark states made from a coloured
SU(3) {\bf 3} combination of $qq$ and a ${\bf \bar 3}$ combination of
$\bar q \bar q$.
This naturally leads to a light $\sigma$, an intermediate $\kappa$
and the highest (degenerate) masses for $a_0(980)$ and $f_0(980)$, in
agreement with experiment.
Note, however that meson-meson
configurations lead to a similar spectrum except that the $a_0(980)$
might lie at the $\eta \pi$ threshold.
This does not happen because of
the nearby Adler zero at $s = m^2_\eta - m^2_\pi /2$.
The $a_0(980)$ migrates to the $KK$ threshold because the Adler zero
in this case is distant, at $ s = m^2_K /2$ \cite {eef}.
Jaffe's  model does not agree
well with the observed decay branching ratio $(\sigma \to KK)/(\sigma
\to \pi \pi)$ near 1 GeV \cite {Recon}.
A serious problem is that, from the width of the $\sigma$ pole, the
$\kappa$ width  is predicted to be $(236 \pm 39)$ MeV, much less than
the latest value: $(758 \pm 10(stat) \pm 44 (syst)$ MeV \cite {kappa}.

A further point is that Maiani et al. \cite {Maiani} extend Jaffe's
scheme to $[c\bar q][\bar qq]$ configurations $\bf {6} \otimes \bar {\bf 3}$.
They give a firm prediction for the observation of analogues of
$a_0(980)$ in $c\bar s n\bar n$ with $I=1$ and charges 0, +1 and
+2.
There is no evidence for such states as yet.

The $\sigma$ and $\kappa$ poles are well predicted in  both
mass and width by the Roy equations \cite {Caprini} \cite {Descotes},
which are based on $t$ and $u$-channel exchanges.
Exchange of $\rho (770)$ and $K^*(890)$ make strong contributions.
The Julich group of Janssen et al. \cite {Janssen} showed that meson
exchanges account for $f_0(980)$ and $a_0(980)$.
It seems unavoidable that all four states $\sigma$, $\kappa$,
$a_0(980)$ and $f_0(980)$ are strongly driven by
meson exchanges.
The Roy equations do however impose the Adler zero coming from
chiral symmetry breaking, a short-range effect.

A final comment concerns what happens to the upper energy combinations
appearing in Eq. (2).
As Jaffe remarks, they become broad and are likely to fall apart,
creating a broad high mass background.
The high mass tail of the $\sigma$ does behave in this sort of way
above 1 GeV, due to coupling to $4\pi$ \cite {f01370}; the precise
form of this high mass behaviour is poorly known because of lack of
data on $\pi \pi \to 4\pi$.
For $J^P = 0^-$, there is also a conspicuous slowly varying component
which appears in data on $J/\Psi \to \gamma 4\pi$ \cite {g4pi};
an alternative explanation in that case is that the broad component 
originates from $J/\Psi \to \gamma GG$, where $G$ are gluons which 
couple strongly to $\rho \rho$ with $L=1$.

The conclusion from this review (and what follows in Section 2)
is that meson loops play a strong role as well as confined
$q\bar q$ and $qqq$ states.
There is as yet no conclusive evidence for anything going beyond these
two components, although there might be small perturbations from
diquarks.
This scenario will now be assumed and confronted with data.

\section {Applications}
\subsection {$X(3872)$}
It seems generally agreed now that the $X(3872)$ is a linear
combination of $c\bar c$ and $\bar D_0D_0^*$, locked close to the $\bar
D_0 D^*_0$ threshold.
A non-resonant cusp alone is too broad to fit the data.
The coupling to $\bar D D^*$ is weaker than that of $f_0(980)$ to $KK$,
but the shape of the dispersion curve is similar; it is evident from
the width of the cusp in $Re \,\Pi$ of Fig. 1 that a cusp alone fails
to fit the $\sim 3$ MeV width of $X(3872)$.
Lee et al. \cite {Lee} make a detailed fit to existing data, solving
the Bethe-Salpeter equation - equivalent to evaluating the dispersion
integral of Eq.  (4).
The binding energy of $X(3872)$ arises
essentially from $\bar D D^*$ loop diagrams.
The magnitude of $\pi $ exchange is known from the width of the
decay $D^* \to D\pi$;
other exchanges are modelled.
However, meson exchanges are not the essential source of binding.
They simply need to be attractive, so that $\bar D$ and $D^*$
approach one another near threshold.
Both $\bar D_0 D^*_0$ and $D^+ D^-$ channels contribute, though the
$X(3872)$ appears at the lower threshold.
The binding energy is controlled sensitively by the form factor.

Kalashnikova and Nefediev point out \cite {Kalash} that there is
decisive evidence for the role of the $\bar cc$ component from the
decay width to $\gamma \Psi '$; this cannot be described by a pure
molecular model.
The branching fraction for $B \to KX$ is also too large to be explained
by a molecule alone.

The narrow width of $X(3872)$ arises because its decay modes to
$\pi ^+\pi ^- J/\Psi$, $\omega J/\Psi$ and possibly $\chi (3510)\sigma$
are OZI suppressed, therefore weak.
Its coupling to $\bar D^0 D^{*0}$ over the width of the resonance is
also weak.
However, the coupling to $\bar D D^*$ rises rapidly above threshold
and produces the binding via virtual loop diagrams.
Kalashnikova and Nefediev conclude that the $\bar cc$ state is
attracted to the $\bar D D^*$ threshold.
Ortega et al. \cite {Ortega} reach a similar conclusion that $X(3872)$
must have a large $\bar D D^*$ component.

Gamermann et al. \cite {Nieves} conclude that a molecular state is
consistent with the decays to $\pi ^+\pi ^- J/\Psi$ and $\omega J/\Psi$,
but do not consider the other evidence for a $\bar cc$ component.

\subsection {Not all cusps are resonances}
There is a cusp at the $\pi d$ threshold \cite {Measday}, but no
resonance.
The exotic $Z^+(4430)$ of Belle \cite {Z4430} is at the
threshold for $D^*(2007) D_1(2420)$ and has a width close to that of
$D_1(2420)$.
The data can be fitted as a resonance, but can also be
fitted successfully by a non-resonant cusp, see Fig. 6 of \cite {Sync}.
Additionally, Babar do not confirm the existence of the $Z(4430)$.

\subsection {Light Vector Mesons}
As an introduction to the vector mesons amongst light quarks, it is
useful to read my review of Crystal Barrel data in flight
\cite {Review}.
The first three figures of this paper show plots of resonance masses
as a function of $s = $ mass squared.
Fig. 2 below shows two examples.
They resemble Regge trajectories, except they are drawn for one set of
quantum numbers at a time.
There is a striking agreement between all trajectories, with a common slope
of $1.143 \pm 0.013$ GeV$^2$ for each unit of excitation.
A similar regularity is observed for baryon resonances with
similar slope \cite {Klempt}.

The $\rho (1900)$ lies close to the $N\bar N$ threshold and it is
well known that the $\bar pp \, ^3S_1$ interaction is strongly
attractive.
It is natural to interpret $\rho (1900)$ as the $n=3$
$^3S_1$ $n\bar n$ state mixed with $\bar pp$, i.e. attracted to the
threshold.
Then other $\rho$ states fall into place as follows:
\newline (ii) $\rho (2000) = ^3D_1,\, n=2$.
It is observed in three sets of data:
$\pi ^+\pi ^-$, $\pi \omega$ and $a_0\omega$.
There are extensive differential cross section and polarisation data on
$\bar pp \to \pi ^+\pi ^-$ from the PS 172 experiment down to a mass of
1910 MeV (a beam momentum of 360 MeV/c) \cite {172}.
There are further similar data above a beam momentum of 1 GeV/c from an
experiment at the Cern PS of Eisenhandler {\it et al} \cite
{Eisenhandler}. The polarisation  data determine the ratio of decay
amplitudes to $^3D_1$ and $^3S_1$  $\bar pp$ configurations:
$r_{D/S}=g_{\bar pp}(^3D_1)/ g_{\bar pp}(^3S_1)=0.70 \pm 0.32$;
for the low available momentum in $\bar pp$,
this is a rather large $^3D_1$ component.
\newline (iii) $\rho (2150) = ^3S_1,\, n=4$.
It is seen in $\pi ^+ \pi ^-$ data of \cite {172} and
\cite {Eisenhandler} and in  Crystal Barrel data for $a_0(980)\pi$ and
in GAMS and Babar data \cite {PDG}; (the Particle Data Group incorrectly 
lists the 1988 MeV state of Hasan \cite {Hasan} under $\rho (2150)$, but it is the
$\rho (2000)$).
For $\rho (2150)$, $r_{D/S}=-0.05 \pm 0.42$.
\newline (iv) $\rho (2265) = ^3D_1$, $n=3$.
It is observed only in two sets of data, $\pi ^+ \pi ^-$ and in Crystal
Barrel data for $a_2\omega$ and therefore needs confirmation; it has a
large error for $r_{D/S}$.
The $\rho (1700)$, $\rho (2000)$ and $\rho (2265)$ are consistent within
errors with a straight trajectory with the same slope as other states,
see Fig. 2(b).
\begin{figure}[htb]
\begin{center} \vskip -12mm
\epsfig{file=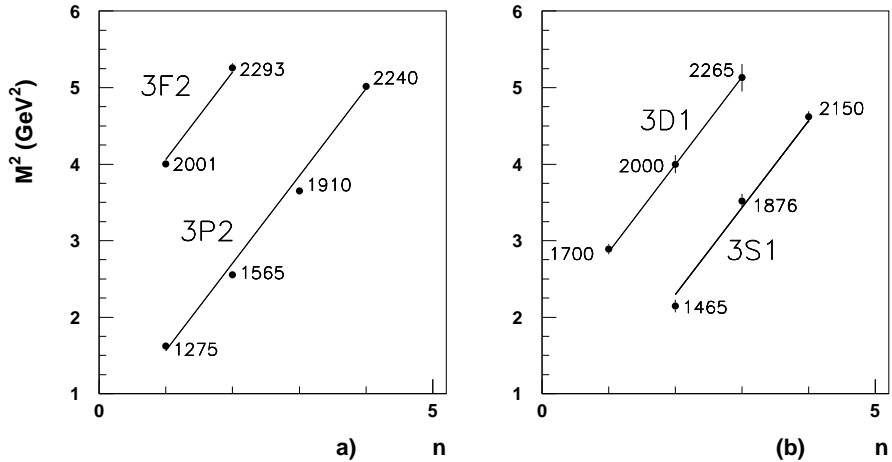,width=14cm}
\vskip -6mm
\caption
{Trajectories of (a) $I=0$, $J^{PC}=2^{++}$ and (b) 
$I=1$, $J^{PC}=1^{--}$ $n\bar n$ resonances.}
 \end{center}
\end{figure}

Confirmation of these states and clarification of the mass range below
1700 MeV would be welcome, particularly the separation into $^3D_1$ and
$^3S_1$.
This could be done at VEPP 2 ($e^+e^-$ up to 2 GeV/c) and VEPP 4
(2--4 GeV/c) in Novosibirsk using large acceptance spectrometers
available there.
If an electron beam with transverse polarisation is used, the initial
state is a 50:50 combination of $^3S_1$ and $^3D_1$, since the
electrons are highly relativistic.
Decays to the final state are not highly relativistic and $^3D_1$
decays contain a conspicuous dependence on the azimuthal angle $\phi$
with respect to the plane defined by the beam and initial polarisation.
Both $\cos 2\phi$ and $\sin \phi$ terms appear, and should provide a
clean separation of the magnitude and phase of the $^3D_1$ decay
amplitude.
\newline (v) The $Y(2175)$ \cite {PDG} observed by BES 2 and Babar
in $\phi f_0(980)$ and $K^+K^- f_0(980)$ makes a natural $s\bar s$
partner for $\rho (2000)$.
Note that there is sufficient momentum in the final state to allow a
$^3D_1$ state.

\subsection {$J^P = 2^+$ light mesons}
The $f_2(1565)$ lies at the $\omega \omega$ threshold.
It is distinctly lower in mass than $a_2(1700)$ and has clearly been
attracted to the $\omega \omega$ threshold.
The mass shift approaches 150 MeV and demonstrates the importance
of the molecular component.
The $f_2(1565)$ also appears clearly in $\pi \pi$, as observed by
several groups \cite {PDG}.
It should appear in $\rho \rho$ with $g^2_{\rho \rho} = 3g^2_{\omega
\omega}$ by SU(2) symmetry, which predicts $g(\rho ^0\rho ^0) = -
g(\omega \omega)$ because of the similar masses of light quarks
and the close masses of $\rho(770)$ and $\omega (782)$.

The $f_2(1640) \to \omega \omega$ observed by GAMS and VES \cite
{PDG} may be fitted by folding the line-shape of $f_2(1565)$ with
$\omega \omega$ phase space and a reasonable form factor \cite {Baker},
together with the dispersive term $Re \, \Pi(s)$.
There is no need for separate $f_2(1565)$ and $f_2(1640)$.
This has confused a number of theoretical predictions of the sequence
of $2^+$ states.

Fig. 2(a) shows trajectories for $2^{++}$ states, including those above
the $\bar pp$ threshold from Crystal Barrel data in flight, using
trajectories with a slope of 1.14 GeV$^2$.

The PDG makes errors in reporting the Crystal Barrel
publications.
It lists $f_2(2240)$ under $f_2(2300)$, which is observed in $\phi
\phi$ and $KK$ by all other groups.
The $f_2(2300)$ is naturally interpreted as an $s\bar s$ state.
Both $f_2(2240)$ and $f_2(2293)$  are observed in a combined
analysis of ten sets of data: four sets  of PS172 and Eisenhandler et
al., together with Crystal Barrel data for $\eta \pi ^0 \pi ^0$, $\eta
'\pi ^0 \pi ^0$, $\eta \eta \eta$, $\pi ^0\pi ^0$, $\eta \eta$ and
$\eta \eta '$.
The data from the last 3 channels are fitted to a linear
combination $\cos \phi \, |n\bar n> + \sin \phi \, |s\bar s>$ and the
mixing angle is determined to be $\phi = 7.5^\circ$ for $f_0(2240)$ and
$\phi = -14.8^\circ$ for $f_2(2293)$ \cite {mixing}. So the $f_2(2240)$
is certainly not an $s\bar s$ state. The $f_2(2240)$ is dominantly
$^3P_2$ with $r_{F/P} = 0.46 \pm 0.09$ (defined like $r_{D/S}$) and the
$f_2(2300)$ is largely $^3F_2$ with $r_{F/P} = -2.2 \pm 0.6$.
The PDG fails to list the $f_2(2293)$ at all.

A further comment on PDG listings is that the $f_2(2150)$ is
conspicuous by its absence from Crystal Barrel data in flight.
All $s\bar s$ states such as $f_2(1525)$ are produced very
weakly in $\bar pp$ interactions.
The $f_2(2150)$ is observed elsewhere mostly in $K\bar K$ and $\eta \eta$
channels.
It is therefore naturally interpreted as an $s\bar s$ state, the
partner of $f_2(1905)$.
The observation of an $f_2(2135)$ by Adomeit et al. \cite {Adomeit}
was qualified in the publication by the warning that it
could be due to the overlap of two or more $f_2$ states.
That was subsequently  shown to be the case \cite {1860} in a later
analysis of data with statistics a factor 7 larger and at 9 momenta
instead of the two used by Adomeit et al.  The later
publication withdrew the claim for an $f_2(2135)$ and the PDG was
informed.

An important systematic observation is that $\bar pp$ states tend to
decay with the same $L$ as the initial $\bar pp$ state.
There is a simple explanation, namely good overlap of the initial
and final states in impact parameter.
This observation may be useful to those calculating decays, hence
mesonic contributions to eqns. (1)--(4).

\subsection {Light $0^+$ mesons}
In BES 2 data for $J/\Psi \to \omega K^+K^-$, there is a clear
$f_0(1710) \to KK$ \cite {WKK}.
In high statistics data for $J/\Psi \to \omega \pi ^+\pi ^-$, \cite
{WPP} there is no visible $f_0(1710)$, setting a limit on branching
ratios:  $BR(f_0(1710) \to \pi \pi )/BR (f_0(1710) \to KK) <0.11$ with
95\% confidence.
Thirdly, in $J/\Psi \to \phi \pi ^+\pi ^-$, there is a
$\pi \pi$ peak requiring an {\it additional} $f_0(1790)$ decaying to
$\pi \pi$ but weakly to $KK$ \cite {phipp}. 
It is readily explained as the radial excitation of $f_0(1370)$. 
There is ample independent
evidence for it in $J/\Psi \to \gamma 4\pi$ \cite {Mark3} \cite {Bai}
\cite {g4pi} and $\bar pp \to \eta \eta \pi ^0$ in flight \cite {eepi}.
BES 2 also report an $\omega \phi$ peak of
95 events at 1812 MeV; $J^P = 0^+$ is favoured \cite {omegaphi}.
It is confirmed by VES data at the Hadron09 conference 
\cite {Ivashin}. 
The BES 1812 MeV peak may be fitted well with the $f_0(1790)$ 
line-shape folded with $\omega \phi$ phase space and a form factor
$\exp -3k^2_{\omega \phi}$.
The exotic signal is naturally interpreted as a $0^+$ glueball
component overlapping this mass range and mixing into $f_0(1790)$:
$$ gg \to (u\bar u + d \bar d + s\bar s)(u\bar u + d \bar d + s\bar s)
\to 4 s\bar s n\bar n$$.

A comment is needed on $f_2(1810)$ of the Particle Data Tables.
It does not fit in naturally in Fig. 2(b).
The spin analysis of the GAMS group \cite {Alde} finds a very marginal
difference between spin 0 and spin 2.
It depends on a fine distinction based on a difference in angular
distibution depending strongly on experimental acceptance; however, no
Monte Carlo of the acceptance is shown.
With the benefit of hindsight, it seems possible that this was in fact
the first observation of $f_0(1790)$.

\subsection {Glueballs}
Morningstar and Peardon \cite {glueball}
predict glueball masses in the quenched approximation where $q\bar q$
are omitted.
When mixing with $q\bar q$ is included, mixing is likely to lower
glueball masses.

\subsection {Broad Thresholds}
Broad thresholds do play a role. For $f_0(1370)$, including
dispersive effects, the Argand diagram still follows a
circle closely.
Experimentalists can then safely omit dispersive effects for such cases
as a first approximation.
However, phase shifts  depart significantly from a Breit-Wigner of
constant width \cite {f01370};
at low mass, where $4\pi$ phase space is small, the phase shift varies
more rapidly and at high mass more slowly.
A further point is that
it is vital to include the effect of rapid changes in phase space
in the Breit-Wigner numerator.
The peak of $f_0(1370) \to 4\pi$ is $\sim 110$ MeV higher than that in
$f_0(1370) \to 2\pi$ because of rapidly increasing $4\pi$ phase space.
Likewise $\eta (1405)$ and $\eta (1475)$ may be fitted as two decay
modes of a single $\eta (1440)$ \cite {1440}.
The $\eta (1475)$ is seen only in $KK^*(890)$, where phase space
rises from threshold near 1385 MeV as
momentum cubed in the final state; the $\eta (1405)$ is seen in
$\kappa K$ S-waves and $\eta \pi \pi$ where phase space changes slowly.

\section {Remarks on further experiments}
There has been important progress in $c\bar c$ spectroscopy of narrow
states and low mass broad states.
This will probably continue for $b\bar b$ states.
However, the chances of sorting out the large
number of broad states at high mass look remote.

Further progress towards a complete spectroscopy of light mesons and
baryons is important for an understanding of confinement - one of the
key phase transitions in physics.
Progress is possible by measuring transverse polarisation in formation
processes.
Consider $\bar pp$ as an example.
The high spin states appear clearly as peaks, e.g. $f_4(2050)$ and
$f_4(2300)$.
These serve as interferometers for lower states.
However, differential cross sections measure only real parts of
interferences.
This leaves the door open to two-fold ambiguities in relative phases
and large errors if resonances happen to be orthogonal.
A measurment of transverse polarisation normal to the plane of
scattering measures $Tr <A^*\sigma _yA>$,
where $A$ is the amplitude. 
This measures the imaginary part of interferences.
The phase sensitivity is important in eliminating ambiguities between
amplitudes.
What appears to be less well known is that transverse polarisation in
the sideways (S) durection gives additional information for three and
four-body final states with a decay plane different to the plane
defined by the beam and initial state polarisation.
This depends on $Tr <A^*\sigma _xA>$, and measures the real parts of
exactly the same interferences as appear from the $\sigma_y$ operator.
Longitudinal polarisation depends only on differences of two
intensities and is less useful.

An example of a simple experiment which would pay a rich dividend
is to measure such polarisations with the Crystal Barrel detector
at the forthcoming GSI $\bar p$ source, over the same mass range as
used at LEAR.
An extracted beam with these momenta will be available at the FLAIR
ring.
Such measurements could indeed have been made at LEAR if it had not
been sacrificed to the funding of the LHC.
The present situation is that the amplitudes for $I=0$, $C=+1$ states
are unique for all expected $J^P$.
For $I=1$, $C=-1$, they are nearly complete, but there are some
weaknesses for low spin states, notably $^3S_1$ which leads to a flat
decay angular distribution.
For $I=1$, $C=+1$ there is a two-fold ambiguity for $\eta \pi$ final
states and crucial $J^P = 0^+$ states are missing.
For $I=0$, $C=-1$ there are many missing states.

The final states to be measured are  (i) $\eta \pi$ and
$\eta \eta \pi$ ($I=1$, $C = +1$), (ii) $\omega \pi$ ($I=1$, $C=-1$),
(iii) $\omega \eta$ and $\omega \eta \pi^0$ ($I=0$, $C=-1$).
A measurement of $\eta \pi ^0 \pi ^0$ would also cross-check the existing
solution and provide information on interferences between singlet
and triplet $\bar pp$ states.
All of these channels can be measured simultaneously with the existing
Crystal Barrel detector.

A Monte Carlo simulation of results extrapolated from existing analyses
predicts a unique set of amplitudes for all quantum numbers.
Data are required from 2 GeV/c down to the lowest possible momentum
$\sim 360$ MeV/c.
This cannot be done by the PANDA experiment which will not go
significantly below 2 GeV/c; the two experiments should be viewed as
complementary.
Seven of the nine momenta studied at LEAR were run in 3 months of 
beam-time, so it is not a long experiment, nor does it demand beam
intensities above $10^5 \bar p/s$.
A Monte Carlo study shows that backgrounds from heavy nuclei in the
polarised target (and its cryostat) should be at or below an average
level of 10\%;
this is comparable with cross-talk between final states and is easily
measured from a dummy target.
A frozen spin target is required, but the technology already exists in
Bonn.
The essential cost is to move the detector to and from Bonn.

Baryon spectroscopy would also benefit from similar $\pi ^\pm p$
measurements, although they are more difficult.
The essential problem is incompatibility between the uniform holding
field required for the polarised target and the use of a magnetic field
over the detector to measure charged particles.
The feasibility of such an experiment is discussed in Ref. \cite
{Experiments}, using a Monte Carlo simulation based on the geometry and
performance of the Crystal Barrel detector.
The CLEO C detector, now idle, though still buried under concrete
shielding, is an obvious alternative.
For $\pi ^-p$, all neutral final states can be measured with
3C kinematics if the neutron is detected in the calorimeter.
For $\pi ^+p$, several important final states are accessible with
2C kinematics, e.g. $N(\pi \pi)$ and $\Delta (\pi\pi )$.
Rates are enormous, so running time is governed essentially by
down-time required for polarising the target and changing momenta.
Data at 30 MeV steps of mass appear sufficient, except close to
2-body thresholds such as $\omega N$.

\section {Conclusions}
The objective of this paper has been to make a case for what appears
logically necessary, namely that both quark combinations at short
range and decay channels at large range contribute to the
eigenstates.
The $X(3872)$ is a prime example of mixing between $c\bar c$ and
meson-meson in the form of $\bar D D^*$.
In view of the calculations of Oset et al. and Barnes and Swanson, it
seems likely that many resonances contain large mesonic contributions.

The data on $a_0(980)$, $f_0(980)$, $f_2(1565)$ and $\rho(1900)$
fit naturally into this picture.
There must be a large mesonic contribution to the nonet of $\sigma$,
$\kappa$, $a_0(980)$ and $f_0(980)$, but there could be a modest
diquark component as well.
It is likely that there will be a small
$q\bar q$ component, but this is suppressed by the $L=1$ centrifugal
barrier for $^3P_0$ combinations; it appears strongly in the mass
range of $f_0(1370)$ and higher states.

Experimentalists must take care to fit the $s$-dependence of the
numerator of Breit-Wigner resonances due to phase space,
e.g. $f_0(1370) \to 2\pi$ has a very different line-shape to
$f_0(1370) \to 4\pi$.
The denominator may be fitted as a first approximation with a 
Breit-Wigner resonance of constant width; however, for high quality data,
the effect of the dispersive component in the real part of the
denominator matters.
For sharp thresholds, e.g. $f_0 \to KK$, the Flatt\' e formula is an
approximation; with high quality data, the correction due to the cusp
in Re $\Pi(s)$ is important, but requires precise information on
experimental resolution.

Further experiments on transverse polarisation in inelastic processes
are needed and appear to be practicable without large cost.

\end{document}